\documentclass[10pt,journal]{IEEEtran}
\makeatletter
\def\ps@headings{%
\def\@oddhead{\mbox{}\scriptsize\rightmark \hfil \thepage}%
\def\@evenhead{\scriptsize\thepage \hfil \leftmark\mbox{}}%
\def\@oddfoot{}%
\def\@evenfoot{}}
\makeatother
\usepackage{amssymb}
\usepackage{amsfonts,amsmath}
\usepackage[ruled]{algorithm}
\usepackage{algpseudocode}
\usepackage{algorithmicx}
\usepackage{graphicx,epsfig,subfigure}
\usepackage{subfig}
\usepackage{color}
\usepackage[table,xcdraw]{xcolor}
\usepackage{booktabs}

\usepackage{cite}

\newcommand {\mymarginpar}[1]{\marginpar{#1}}
\renewcommand {\marginpar}[1]{}

\def\_{\rule{.3em}{.15ex}}      

\newcommand{\ls}[1]
   {\dimen0=\fontdimen6\the\font
    \lineskip=#1\dimen0
    \advance\lineskip.5\fontdimen5\the\font
    \advance\lineskip-\dimen0
    \lineskiplimit=.9\lineskip
    \baselineskip=\lineskip
    \advance\baselineskip\dimen0
    \normallineskip\lineskip
    \normallineskiplimit\lineskiplimit
    \normalbaselineskip\baselineskip
    \ignorespaces
   }

\newcommand {\bearn}{\begin{eqnarray*}}
\newcommand {\eearn}{\end{eqnarray*}}
\newcommand {\barr}{\begin{array}}
\newcommand {\earr}{\end{array}}

\newcommand {\N}{{\cal N}}

\newtheorem{definition}{Definition}
\newtheorem{property}[definition]{Property}
\newtheorem{proposition}[definition]{Proposition}
\newtheorem{lemma}[definition]{Lemma}
\newtheorem{theorem}[definition]{Theorem}
\newtheorem{corollary}[definition]{Corollary}
\newtheorem{example}[definition]{Example}
\newtheorem{remark}[definition]{Remark}

\newcommand{\comb}[2]
{\left ( \begin{array}{c} #1 \\#2 \end{array} \right ) }

\newcommand {\benum} {\begin{enumerate}}
\newcommand {\eenum} {\end{enumerate}}

\newcommand {\bdesc} {\begin{description}}
\newcommand {\edesc} {\end{description}}

\newcommand {\bfig}[2] {\begin{figure}
  \centering
  \includegraphics[width=#2]{#1}}
\newcommand {\brotatefig}[2] {\begin{figure}[htbp]
                        \centerline {
                         \epsfig{figure={#1},clip=,angle=-90,width={#2}}}}
\newcommand {\bfigfirst}[2] {\begin{figure}[h]
                        \centerline {
                        \setlength{\epsfxsize}{#2}
                        \epsffile{#1}}}
\newcommand {\efig}[2]{ \caption{#2}
                        \label{fig:#1}
                        \end{figure}
                        \mymarginpar{fig:#1}}
\newcommand {\erotatefig}[2]{ \caption{#2}
                        \label{fig:#1}
                        \end{figure}
                        \mymarginpar{fig:#1}}
\newcommand {\rfig}[1]{Figure \ref{fig:#1}}

\newcommand {\btab}[1]{
                       \begin{table}
                       \centering
                       \begin{tabular}{#1}}
\newcommand {\etab}[3] {
                       \end{tabular}
                       \caption[#3]{#2}
                       \label{tab:#1}
                       \end{table}
                       \mymarginpar{tab:#1}
                       \vspace{.1in}}

\newcommand {\btabular}[1]{\begin{center}
                       \begin{tabular}{#1}}
\newcommand {\etabular}{\end{tabular}
                       \end{center}}

\newcommand {\bdefin}[1]{\begin{definition}
                      \mymarginpar{def:#1}
                      \label{def:#1} }
\newcommand {\edefin}       {\end{definition}}
\newcommand {\rdef}[1]{Definition \ref{def:#1}}

\newcommand {\bpro}[1]{\begin{property}
                      \mymarginpar{pro:#1}
                      \label{pro:#1} }
\newcommand {\epro}   {\end{property}}

\newcommand {\bprop}[1]{\begin{proposition}
                      \mymarginpar{prop:#1}
                      \label{prop:#1} }
\newcommand {\eprop}       {\end{proposition}}
\newcommand {\rprop}[1]{Proposition \ref{prop:#1}}

\newcommand {\blem}[1]{\begin{lemma}
                      \mymarginpar{lem:#1}
                      \label{lem:#1} }
\newcommand {\elem}   {\end{lemma}}
\newcommand {\rlem}[1]{Lemma \ref{lem:#1}}

\newcommand {\bthe}[1]{\begin{theorem}
                      \mymarginpar{the:#1}
                      \label{the:#1} }
\newcommand {\ethe}   {\end{theorem}}
\newcommand {\rthe}[1]{Theorem \ref{the:#1}}

\newcommand {\bproof}{\noindent {\bf Proof.} \ }
\newcommand {\eproof} {\hfill \squares \\ \vspace{.3cm}}
\newcommand {\bcor}[1]{\begin{corollary}
                      \mymarginpar{cor:#1}
                      \label{cor:#1} }
\newcommand {\ecor}   {\end{corollary}}

\newcommand {\bax}[1]{\begin{axiom}
                      \mymarginpar{ax:#1}
                      \label{ax:#1} }
\newcommand {\eax}       {\vspace{-.1in} \end{axiom}}

\newcommand {\bex}[2]{\vspace{.1in}
                      \begin{example}
                      \mymarginpar{ex:#1}
                       {\bf #2}
                      \label{ex:#1} }
\newcommand {\eex}       {\end{example} \vspace{.3cm} }

\newcommand {\brem}[1]{\begin{remark}
                      \mymarginpar{rem:#1}
                      \label{rem:#1} \em }
\newcommand {\erem}   {\end{remark}}

\newcommand {\beq}[1]{\mymarginpar{eq:#1}
                      \begin{equation}
                      \label{eq:#1} }

\newcommand {\beqno}[1]{\mymarginpar{eq:#1}
                      \begin{eqnarray}
                      \nonumber}

\newcommand {\eeq}       {\end{equation}}
\newcommand {\eeqno}       { && \end{eqnarray}}
\newcommand {\req}[1]{(\ref{eq:#1})}

\newcommand {\bear}[1]{\mymarginpar{eq:#1}
                       \begin{eqnarray}
                       \label{eq:#1} }

\newcommand {\bearno}[1]{\mymarginpar{eq:#1}
                       \begin{eqnarray}
                       \nonumber}

\newcommand {\eear}{\end{eqnarray}}
\newcommand {\eearno}{\end{eqnarray}}
\newcommand {\bsel}{\left \{ \begin{array}{cl}}
\newcommand {\esel}{\end{array} \right.}

\newcommand {\bmat}[1]{\left [ \begin{array}{#1}}
\newcommand {\emat}{\end{array} \right ]}
\newcommand {\bsec}[2]{\mymarginpar{sec:#2}
                       \section{#1}
                       \label{sec:#2} }

\newcommand {\bsubsec}[2]{\mymarginpar{sec:#2}
                       \subsection{#1}
                       \label{sec:#2} }

\def\R{I\kern-0.30em R}
\def\N{I\kern-0.30em N}
\def\P{I\kern-0.30em P}
\newcommand\squares{\vrule height6pt width7pt depth1pt}

\def\ex{{\bf\sf E}}
\def\pr{{\bf\sf P}}

\newcommand{\WW}{b}
\newcommand{\MM}{M}
\newcommand{\uu}{u}

\begin{document}

\title{A Hierarchical Stitching Algorithm for Coded Compressed Sensing}

\author{Yi-Jheng Lin, Chia-Ming Chang, and Cheng-Shang~Chang,~\IEEEmembership{Fellow,~IEEE},
\thanks{Y.-J. Lin, C.-M. Chang, and C.-S. Chang are with  the Institute of Communications Engineering,
National Tsing Hua University,
Hsinchu 300, Taiwan, R.O.C.
email:   s107064901@m107.nthu.edu.tw, jamie@gapp.nthu.edu.tw, cschang@ee.nthu.edu.tw.}
}
\maketitle

\begin{abstract}
Recently, a novel coded compressed sensing (CCS) approach was proposed in \cite{amalladinne2020coded} for dealing with the scalability problem for large sensing matrices in massive machine-type communications. The approach is to divide the compressed sensing (CS) problem into smaller CS sub-problems. However, such an approach requires stitching the results from the sub-problems to recover the result in the original CS problem. For this stitching problem, we propose a hierarchical stitching algorithm that is easier to implement in hardware for parallelization than the tree coding algorithm in \cite{amalladinne2020coded}. For our algorithm, we also derive an upper bound on the probability of recovery errors.
\end{abstract}




\bsec{Introduction}{introduction}

There are three connectivity services for the fifth-generation networks (5G) and beyond: (i) enhanced mobile broadband (eMBB), (ii) ultra-reliable low-latency communications (URLLC), and (iii) massive machine-type communications (mMTC) (see, e.g., \cite{bennis2018ultra,popovski2019wireless,le2020overview} and references therein).
For mMTC, the aim is to provide communications in the context of the Internet of Things (IoT). As mentioned in \cite{polyanskiy2017perspective}, there
are three interesting aspects 
for mMTC: (i) small size of the payload leads
to finite-blocklength (FBL) effects, (ii) only a small
fraction of users are active at any given time (random access),
 and (iii) users access channel without any
prior resource requests to the base station (grant-free or uncoordinated transmissions).
To cope with these three aspects for mMTC, Ordentlich and Polyanskiy \cite{ordentlich2017low} considered the $T$-user Gaussian multiple access channel (T-GMAC) and proposed a low-complexity coding scheme,
in which the transmission period is divided into synchronous time slots, and every active user transmits a codeword
during a randomly chosen time slot. 
As mentioned in the recent paper \cite{amalladinne2020coded}, there is approximately a 20 dB gap between the
coding scheme in \cite{ordentlich2017low}
and the achievability limit associated with
the unsourced MAC in \cite{polyanskiy2017perspective}. By using successive interference
cancellation (SIC) across time slots as in 
coded slotted ALOHA (CSA) \cite{liva2011graph,narayanan2012iterative,paolini2012random,stefanovic2018coded}
the gap was pushed to 6dB in \cite{vem2019user}.

In \cite{amalladinne2020coded,fengler2019sparcs}, a compressed sensing (CS) approach was proposed for the uncoordinated and unsourced MAC problem. Their approach is to treat  uncoordinated multiple
access with a massive number of users as support recovery in noisy compressed
sensing. Suppose that $B$ bits are transmitted by each user. Then the CS approach requires a sensing matrix with 
$2^B$ columns, and that makes it computationally infeasible if $B$ is on the order of 100.
To address the scalability problem, the idea in \cite{amalladinne2020coded} is to divide the $B$ bits into $n$ sub-blocks so that the  sub-block size of the sub-CS problem is within the computational capability of state-of-the-art technology.
However, this also raises the problem to pierce the $n$ sub-blocks sent by the same user together. For this, a novel coded compressed sensing (CCS) scheme was devised in \cite{amalladinne2020coded}. At the end of each sub-block, parity bits are padded by hashing the previous sub-blocks (of the same user) with random linear codes. A low-complexity tree coding algorithm that iteratively explores the sub-blocks in the next time slot, and adds candidate sub-blocks (that pass the consistency check of parity bits) into a candidate tree of each user. In the $n^{th}$ time slot, if the candidate tree contains only one path, then the codeword of a user can be recovered. Instead of using the hard stitching decisions in \cite{amalladinne2020coded}, a recent advance in \cite{amalladinne2020unsourced}  is to use soft stitching decisions based on LDPC codes (with parity sub-blocks) so that the sub-CS problem and the stitching problem can be iteratively solved.
However, as pointed out in \cite{ebert2021stochastic}, there might be several codewords on the same factor
graph within the LDPC code, and  that leads to negative impacts for 
the application of belief propagation on the LDPC code.

The exploration of the tree coding algorithm in \cite{amalladinne2020coded} is {\em sequential} from one time slot to another time slot. There are several disadvantages of that: (i) (Parallelization) it is not easy to {\em parallelize} the tree decoding as the candidate tree in one time slot depends on the candidate tree in the previous time slot, (ii) (Identical sub-blocks) the parity bits only depend on the information of the previous sub-blocks, and it might be problematic when two users have identical sub-blocks in early time slots, (iii) (Hardware implementation) to balance the growth of the candidate tree, the number of parity bits need to be tuned and
might be different in every time slot. Variable sub-block formats in the $n$ time slots complicate the hardware implementation of the algorithm.

To facilitate the hardware design for parallelization, in the letter we propose a hierarchical stitching algorithm with an identical sub-block format in every time slot. For our algorithm, we require that the number of time slots $n$ is a power of 2 so that the $n$ sub-blocks can be ``merged'' in a complete binary tree.
Instead of exploring candidates for one user from one time slot to another time slot, we explore candidates for all the users in two adjacent time slots, and that leads to $n/2$ candidate sets that contain fragments in chains of two sub-blocks.
Then we recursively move up one level of the complete binary tree to double the length of fragments.
When the root of the binary tree is reached, we recover all the messages sent by all the users.

We summarize the contributions of this letter for the coded compressed sensing problem below.

\noindent (i) A hierarchical stitching algorithm with a fixed sub-block format   that enables parallelization in hardware.

\noindent (ii) A mathematical analysis for the expected number of erroneous candidates in the hierarchical stitching algorithm that leads to an upper bound for the probability of recovery errors. Such a bound holds even when there are identical sub-blocks from two different users.

\noindent (iii)  An allocation scheme for the number of parity bits at each level of the binary tree that can be used for controlling the computational complexity.

\bsec{The stitching problem}{stitch}

In the letter, we consider the coded compressed sensing scheme for unsourced multiple access in \cite{amalladinne2020coded}. As mentioned in the introduction, the approach in \cite{amalladinne2020coded}
consists of two steps: (i) decomposing the CS problem into several CS sub-problems, and (ii) stitching the results from
sub-problems to recover the results in the original problem. In this letter, 
we only consider the stitching problem. Suppose that there are $K$ users. Let
${\cal L}= \{\underline v_1, \underline v_2, \ldots, \underline v_K\}$ be the collection of $K$ codewords sent by the $K$ users.
Each codeword is divided into $n$ sub-blocks and it has the following binary representation:
\beq{code1111}
\underline v_i=\underline w_i(0)\underline p_i(0)\underline w_i(1)\underline p_i(1) \cdots \underline w_i(n-1)\underline p_i(n-1),
\eeq
where $\underline w_i(j)$ represents the information bits in the $j^{th}$ sub-block of user $i$ and
$\underline p_i(j)$ represents the parity bits in the $j^{th}$ sub-block of user $i$.
The $K$ sub-blocks $\{\underline w_i(j)\underline p_i(j), i=0,1, \ldots, K-1\}$, are received at the receiver in the $j^{th}$ time slot, $j=0,1,\ldots, n-1$. However, the receiver does not know which block belongs to which codeword (for unsourced MAC). The stitching problem is to use the parity bits to perform a consistency check and  then stitch the $n$ sub-blocks of the same codeword together.

As there are $K$ codewords among $K^{n-1}$ ways of stitching sub-blocks, we have the following lower bound for the stitching problem.

\bprop{lbound}
Suppose that there are exactly $K$ codewords for the stitching problem with $n$ time slots. Then the number of parity bits needed for correct recovery of the $K$ codewords (up to a permutation) is at least 
\bear{lbound1111}
&&\log_2  \comb {K^{n-1}} {K} =\log_2 \Big (\prod_{k=0}^{K-1}\frac{(K^{n-1}-k)}{K-k} \Big)\nonumber\\
&& \ge (n-2)K \log_2K.
\eear
\eprop

\bsec{Hierarchical stitching}{hier}

In this section, we propose the hierarchical stitching algorithm. Our algorithm is basically a divide-and-conquer algorithm.
Assume that the number of sub-blocks $n$ is a power of 2, i.e., $n=2^\MM$ for some $\MM$.
The key idea is to  stitch the $2\ell^{th}$ sub-block and $(2\ell+1)^{th}$ sub-block into a chain of two sub-blocks, for $\ell=0,1, \ldots 2^{\MM-1}-1$. By viewing every chain of two sub-blocks as a new sub-block,
we reduce the stitching problem to the size of $2^{\MM-1}$ sub-blocks. By recursively carrying out the stitching operation of two adjacent sub-blocks, we eventually obtain a single chain of $2^\MM$ sub-blocks.

\bsubsec{Complete binary tree}{complete}

\begin{figure}[ht]
	\centering
	\includegraphics[width=0.45\textwidth]{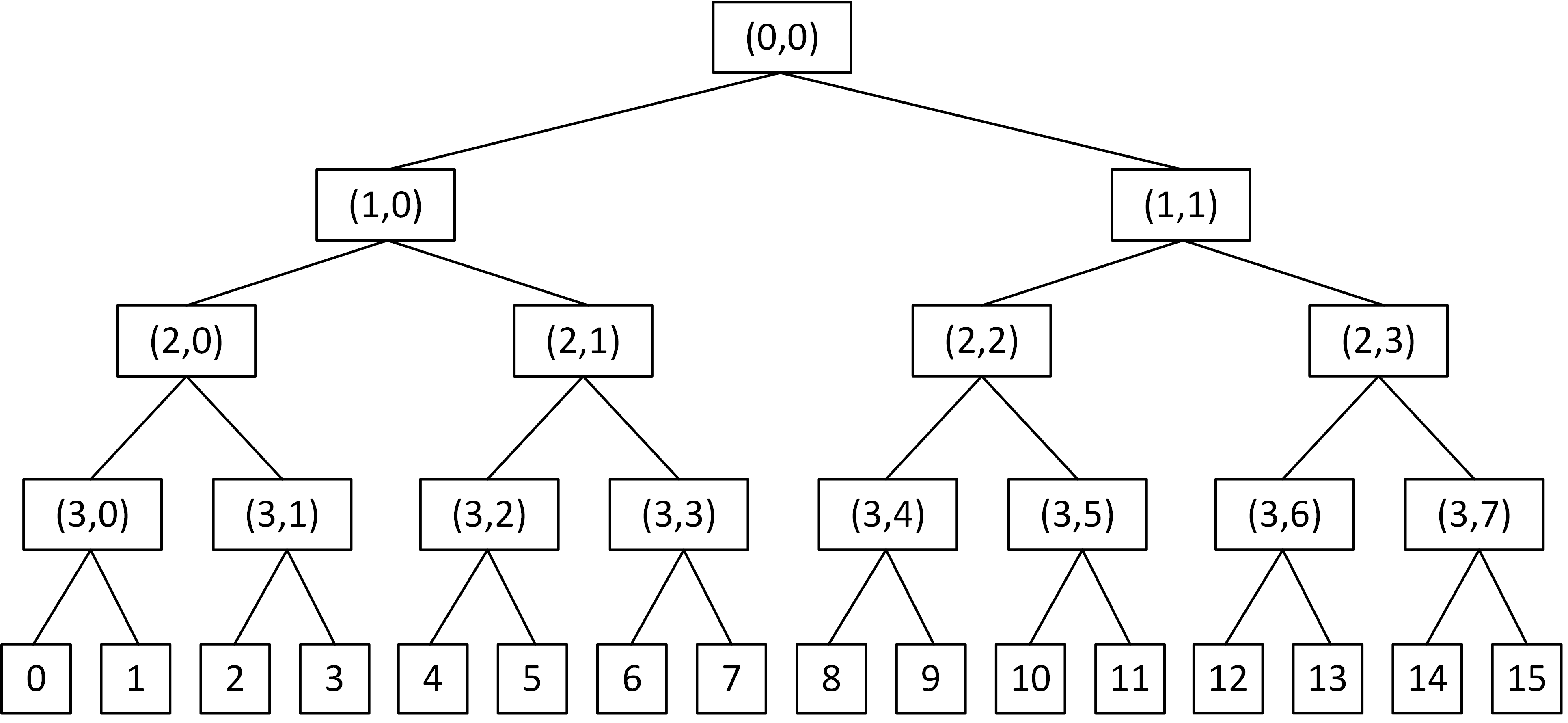}
	\caption{A complete binary tree with 16 leaves (sub-blocks).}
	\label{fig:binarytree}
\end{figure}

To explain our stitching algorithm in detail, let us  consider a complete binary tree with $2^\MM$ leaves as in \cite{chueh2013load}
(see \rfig{binarytree} for a complete binary tree with 16 leaves).
In such a binary tree, there are $\MM+1$ levels, indexed from $0,1,\ldots, \MM$. The root is the only node at level 0, and
a node is at level $m$ if it is a child of a node at level $m-1$.
Index the root as node $(0,0)$, and index recursively the two children of node $(m,\ell)$
as nodes $(m+1,2\ell)$ and $(m+1,2\ell+1)$ for $0\leq \ell\leq 2^m-1$ and $0\leq m\leq \MM-1$.
Clearly, there are $2^m$ nodes at level $m$.
Note that
the $2^\MM$ leaves are simply nodes $(\MM,0), \ldots, (\MM, 2^\MM-1)$ and
 we will simply call node $(\MM,j)$ as leaf $j$ (by omitting the index of level $\MM$).



For $0\leq m\leq \MM$ and $0\leq \ell\leq 2^m-1$,
we now define the $\ell^{th}$ subtree at the $m^{th}$ level as
the subtree constructed by node $(m,\ell)$ and all its descendants.
Such a subtree is called subtree $T(m,\ell)$. Note that
the set of all the leaves in subtree $T(m,\ell)$ is
\begin{eqnarray}
\label{eq:bit-reverse_111}
\{j|\ell \cdot 2^{\MM-m}\leq j\leq (\ell+1)2^{\MM-m}-1\}.
\end{eqnarray}
Thus, the total number of leaves in
subtree $T(m,\ell)$ is $2^{\MM-m}$.

Now view every sub-block as a leaf node in a complete binary tree.
The stitching algorithm starts from merging two adjacent leaf nodes at level $\MM$, and that reduces the number of levels of the complete binary tree by 1. It then recursively merges two adjacent subtrees from level $\MM-1$ to 0.
When the root node is reached, we complete the stitching of the $n$ sub-blocks.

\bsubsec{Parity bits}{parity}

\begin{figure}[ht]
	\centering
	\includegraphics[width=0.4\textwidth]{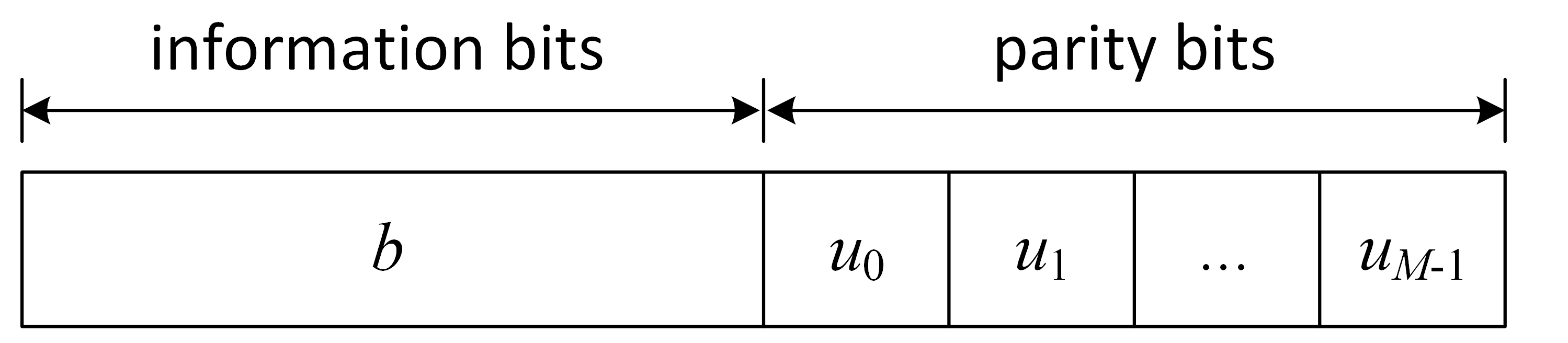}
	\caption{The sub-block format. }
	\label{fig:format}
\end{figure}

In this section, we discuss how parity bits are generated in our algorithm.
To ease  the hardware design, we consider a fixed format for each sub-block.
In each sub-block, there are $\WW$ information bits and $\MM$ levels of parity bits.
The number of level $m$ parity bits is $\uu_m$, $m=0,1,\ldots, \MM-1$  (see \rfig{format} for an illustration of the sub-block format). Note that the total number of parity bits in each sub-block is $\sum_{m=0}^{\MM-1}\uu_m$. Thus,  the total number of information bits (resp. parity bits) in a codeword is $2^\MM \WW$ (resp. $P=2^\MM \sum_{m=0}^{\MM-1}\uu_m$).

The purpose of level $m$ parity bits is for the consistency check when two subtrees at level $m+1$ are merged into a subtree at level $m$.
As each non-leaf node has two subtrees, we call these two subtrees {\em sibling subtrees}.
Specifically, the subtree $T(m,\ell)$ has two sibling subtrees $T(m+1,2\ell)$ and $T(m+1, 2\ell+1)$.
Since there are $2^{\MM-m}$ leaf nodes of $T(m,\ell)$, the total number of information bits
for the sub-blocks in the leaf nodes of $T(m,\ell)$ is $\WW 2^{\MM-m}$.
The idea is to use a hash function to map these  $\WW 2^{\MM-m}$ information bits to $\uu_m 2^{\MM-m}$ bits, and assign them to the $2^{\MM-m}$ leaf nodes in the subtree $T(m,\ell)$
so that each leaf node (sub-block) is assigned  with $\uu_m$ level $m$ parity bits.
For example, consider the binary tree in \rfig{binarytree}. Suppose that $\uu_2=1$. To merge the two sibling subtrees $T(2,0)$ and $T(2,1)$ into $T(1,0)$ in \rfig{binarytree}, the information bits in sub-blocks 0,1,2,3,4,5,6, and 7, are hashed into 8 bits $c(0), c(1),\ldots c(7)$. Then bit $c(j)$ is assigned to the level 2 parity bit of the $j^{th}$ sub-block, $j=0,1,\ldots, 7$.

In Algorithm \ref{alg:parity}, we show the detailed steps to generate the parity bits.

\begin{algorithm}\caption{Generating the parity bits}\label{alg:parity}

\noindent {\bf Input}  $n \WW$ information bits with $n=2^\MM$. The number of parity bits at level $m$, $\uu_m$, $m=0,1,\ldots, \MM-1$, and a set of hash functions $h_m$, $m=0,1,\ldots, \MM-1$.

\noindent {\bf Output} Coded $n$ sub-blocks with parity bits.

\noindent 1: Partition $n\WW$ information bits into $n$ sub-blocks, each with $\WW$ information bits.

\noindent 2: {\em For} $m=\MM-1,\MM-2, \ldots, 0$,

\noindent 3: \quad {\em For} $\ell=0,1, \ldots ,2^{m}-1$,

\noindent 4: \quad \quad Use the hash function $h_m$ to map the  $\WW 2^{\MM-m}$ information bits in the subtree $T(m,\ell)$ into $\uu_m 2^{\MM-m}$ bits.

\noindent 5: \quad \quad Assign the $\uu_m 2^{\MM-m}$ bits to the $2^{\MM-m}$ leaf nodes in the subtree $T(m,\ell)$
so that each leaf node (sub-block) is assigned with $\uu_m$ level $m$ parity bits.

\noindent 6: \quad {\em End}

\noindent 7: {\em End}

\end{algorithm}

\bsubsec{The stitching algorithm}{algorithm}

Suppose that $K$ users are sending 
$K$ codewords to the receiver and that each codeword contains $n=2^\MM$ sub-blocks
with the padded parity bits described in Algorithm \ref{alg:parity}. Index the $n$ time slots from 0 to $2^\MM-1$. 
In the $\ell^{th}$ time slot, the $\ell^{th}$ sub-blocks from the $K$ users are received, but they are mixed together and
the receiver cannot tell which sub-block belongs to which codeword without looking into the parity bits.
Note that two users may send an identical sub-block in the same time slot. As such, the total number of sub-blocks received in a time slot might be less than $K$.
An exhaustive search for the $K$ codewords requires $O(K^{n-1})$ computational complexity.
Our stitching algorithm can greatly reduce the computational complexity by exploiting the power of a divide-and-conquer algorithm. 
The idea is to keep a set of ``candidates'' for each subtree (of the $K$ codewords) during the hierarchical stitching process. The candidate set for each subtree contains ``chained'' sub-blocks that pass the consistency check (for the parity bits) up to the level of the subtree. Initially, for each time slot $\ell$, the candidate set at the leaf (sub-block) level, i.e., level $\MM$, simply contains the $K$ sub-blocks from the $K$ codewords. To move up one level,
we select a candidate from a subtree and another candidate from its sibling subtree, and perform the consistency check by using the hash functions in Algorithm \ref{alg:parity}. If the concatenation of these two chained sub-blocks passes the consistency check, then it is added as a candidate in the parent subtree of these two sibling subtrees. Clearly, the size of each candidate set at level $\MM$ is $K$, and thus it requires $K^2$ consistency check operations to find the candidate set at level $\MM-1$.
The key is to maintain the expected sizes of the candidate sets at level $\MM-1$ within $O(K)$ so that the number of consistency check operations remains $O(K^2)$ when we move up one level. The chaining (merging) process is repeated until the root node (level 0) is reached. If there are no errors, one should have exactly $K$ chained sub-blocks in the candidate set at level 0, and these $K$ chained sub-blocks are the $K$ codewords sent by the $K$ users.
The detailed steps of the hierarchal stitching algorithm are given in Algorithm \ref{alg:stitching}.

\begin{algorithm}\caption{The hierarchical stitching algorithm}\label{alg:stitching}

\noindent {\bf Input}  $K$ sub-blocks (with the format in Algorithm \ref{alg:parity}) in each of the $n=2^\MM$ time slots, and the hash functions $h_m$, $m=0,1,\ldots, \MM-1$, in Algorithm \ref{alg:parity}.

\noindent {\bf Output} The $K$ codewords from the $K$ users.

\noindent 1: Let $S(\MM,\ell)$ be the candidate set that contains at most $K$ sub-blocks at the $\ell^{th}$ time slot, $\ell=0,1,\ldots, 2^\MM-1$.

\noindent 2: Let $S(m,\ell)$ be an empty set that represents the initial candidate set for the subtree $T(m,\ell)$, $m=\MM-1,\MM-2, \ldots, 0$, $\ell=0,1, \ldots ,2^{m}-1$.

\noindent 3: {\em For} $m=\MM-1,\MM-2, \ldots, 0$,

\noindent 4: \quad {\em For} $\ell=0,1, \ldots ,2^{m}-1$,

\noindent 5: \quad \quad {\em For} each pair of two candidates in $S(m+1,2\ell) \times S(m+1,2\ell+1)$,

\noindent 6: \quad \quad \quad Concatenate these two candidates (chained sub-blocks) into a single chained sub-block, and perform the consistency check
by using the hash function $h_m$.

\noindent 7: \quad \quad \quad  If it passes the consistency check, add it to the candidate set $S(m,\ell)$.

\noindent 8: \quad \quad {\em End}

\noindent 9: \quad {\em End}

\noindent 10: {\em End}

\noindent 11: Output the candidate set $S(0,0)$.

\end{algorithm}

\begin{figure}[ht]
	\centering
	\includegraphics[width=0.4\textwidth]{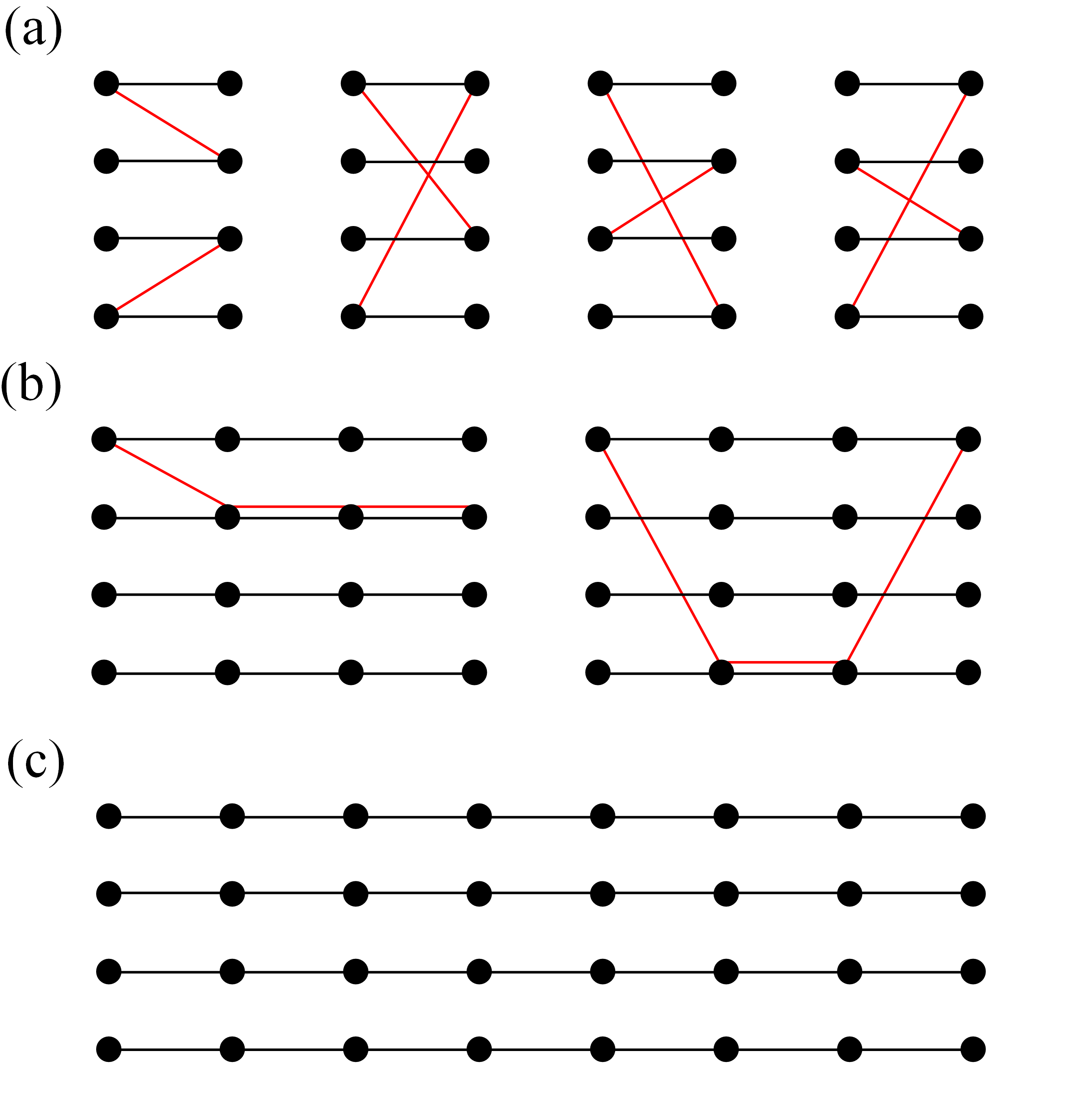}
	\caption{An illustration of the hierarchical stitching algorithm with $K=4$ codewords and $n=8$ sub-blocks: (a) the four candidate sets at level 2, (b) the two candidate sets at level 1, and (c) the candidate set at level 0 (with 4 correct codewords). The black (resp. red) paths represent the correct (resp. erroneous) candidates. }
	\label{fig:stitchingexample}
\end{figure}

In \rfig{stitchingexample}, we provide an illustrating example of the hierarchical stitching algorithm with $K=4$ codewords and $n=8$ sub-blocks. We represent a sub-block as a node in an $n$-stage multi-partite graph. A path in the
multi-partite graph represents a chain of sub-blocks. 
As $\MM=3$, there are three subtree levels: level 2 in \rfig{stitchingexample} (a),
level 1 in \rfig{stitchingexample} (b), and level 0 in \rfig{stitchingexample} (c). As shown in \rfig{stitchingexample},
there are four subtrees at level 2, $T(2,0)$, $T(2,1)$, $T(2,2)$, and $T(2,3)$. The candidate set for $T(2,0)$ (the leftmost bipartite graph) contains 6 candidates with each candidate represented by a chain of two sub-blocks. As $K=4$, there are 4 correct candidates in the candidate set of $T(2,0)$ (marked in black) and two erroneous candidates in the candidate set of $T(2,0)$ (marked in red). At level 1, the two candidate sets in $T(2,0)$ and $T(2,1)$ are merged into the candidate set for $T(1,0)$ (the left 4-partite graph in \rfig{stitchingexample}(b). The candidate set for $T(1,0)$ still contains 4 correct candidates. But there is only one erroneous candidate (with a chain of four sub-blocks).
Finally, at level 0, the two candidate sets for $T(1,0)$ and $T(1,1)$ are merged into the candidate set for $T(0,0)$ and this candidate set only contains the four correct candidates that are the original four codewords sent by the four users.

\bsubsec{Analysis}{analysis}

In this section, we conduct an analysis for Algorithm \ref{alg:stitching}.
A candidate for the subtree $T(m,\ell)$ is a chain of $2^{\MM-m}$ sub-blocks that satisfies the consistency check up to level $m$.
Let $L(m,\ell)$ be the number of candidates for the subtree $T(m,\ell)$ and $L_{\rm err}(m,\ell)$ be  the number of {\em erroneous} candidates for the subtree $T(m,\ell)$.
As in \cite{amalladinne2020coded}, we will show how to bound $\ex [L_{\rm err}(m,\ell)]$ if the hash functions described in the previous section satisfy the {\em uniform hashing assumption} defined below.

\bdefin{SUHA}{(\bf uniform hashing assumption)}  Let $Z_N$ be the set of integers $\{0,1,\ldots, N-1\}$. Consider a family of functions from some set $S$ to $Z_N$.
 A randomly selected  function $h$ in the family of functions is said to  satisfy the {\em uniform hashing assumption} if for any $a \ne b \in S$,
  \beq{suha1111}
 \pr (h(a)=h(b))=\frac{1}{N}.
 \eeq
 \edefin
 
 In Lemma 11 of \cite{amalladinne2020coded}, it was shown that random linear codes can be used for generating parity bits that satisfy the uniform hashing assumption.

\blem{ubound}
Suppose that (A1) the hash functions satisfy the uniform hashing assumption, and (A2) all the information bits are independent Bernoulli random variables with parameter $1/2$.

\noindent (i) For $\ell=0,1, \ldots, 2^{\MM-1}-1$,
\beq{ubound5566}\ex [L_{\rm err}(\MM-1,\ell)] \le {K^2}(\frac{1}{2})^{2 \uu_{\MM-1}}.
\eeq

\noindent (ii)  For $\ell=0,1, \ldots, 2^{\MM-1}-1$,
\beq{ubound5577}
\ex [L(\MM-1,\ell)] \le K+{K^2}(\frac{1}{2})^{2 \uu_{\MM-1}}.
\eeq

\noindent (iii)  For $m=\MM-2,\ldots, 0$, $\ell=0,1, \ldots, 2^{m}-1$,
\bear{ubound1111}
&&\ex [L_{\rm err}(m,\ell)] \le \nonumber\\
&& \ex [L(m+1,2\ell)]\ex [L(m+1,2\ell+1)]{(\frac{1}{2})^{\uu_m 2^{\MM-m}}},\nonumber\\
\eear
and
\bear{ubound2222}
\ex [L(m,\ell)] \le K +\ex [L_{\rm err}(m,\ell)].
\eear
\elem

\bproof
(i)
At level $\MM-1$, an erroneous candidate occurs  when a chain of two sub-blocks passes the consistency check, and it is not a fragment of a codeword. As there are $\uu_{\MM-1}$ parity bits at level $\MM-1$ for each sub-block, to merge two sub-blocks requires the $2\uu_{\MM-1}$ parity bits to be the same as the hashing output of the $2\WW$ information bits in these two sub-blocks.
From  the uniform hashing assumption in \rdef{SUHA} and the independent assumption of information bits, this happens with probability $(\frac{1}{2})^{2 \uu_{\MM-1}}$.
Since there are at most $K^2$ ways of forming an erroneous candidate, the expected number of erroneous candidates is
bounded by ${K^2}(\frac{1}{2})^{2 \uu_{\MM-1}}$.

(ii)  Since there are at most $K$ corrected candidates, the expected number of candidates at level $\MM-1$ is bounded by $K+{K^2}(\frac{1}{2})^{2 \uu_{\MM-1}}$. 

(iii) The argument is basically the same as (i) and (ii). Consider a  candidate in subtree $T(m+1,2\ell)$ and another candidate in subtree $T(m+1,2\ell+1)$. To merge these two candidates requires  the concatenation of the $\uu_m 2^{\MM-m-1}$ parity bits in the subtree $T(m+1,2\ell)$ and the $\uu_m 2^{\MM-m-1}$ parity bits in the subtree $T(m+1,2\ell+1)$
to be the same as the hashing output of the concatenation of the $\WW 2^{\MM-m-1}$ information bits in the  subtree $T(m+1,2\ell)$ and the $\WW 2^{\MM-m-1}$ information bits in the subtree $T(m+1,2\ell+1)$.
Once again,  from the uniform hashing assumption and the independent assumption of information bits, an erroneous candidate at level $m$ happens with probability $(\frac{1}{2})^{\uu_m 2^{\MM-m}}$. Since there are at most $L(m+1,2\ell)L(m+1,2\ell+1)$ ways to form an erroneous candidate at level $m$, we have from the independent assumption that the expected number of erroneous candidates is bounded by
$$(\ex [L(m+1,2\ell)] \ex [L(m+1,2\ell+1)]){(\frac{1}{2})^{\uu_m 2^{\MM-m}}}.$$

Since there are at most $K$ corrected candidates,  the expected number of candidates at level $m$ is then bounded by
\req{ubound1111}.
\eproof

We note that an exact analysis for the expected number of candidates is complicated as two codewords might have the same information bits and parity bits in a chain of two sub-blocks. As such, the number of correct candidates might be less than $K$, and we have to restrict ourselves to the derivation of the upper bounds in \rlem{ubound}.

In the following theorem, we show that if the number of parity bits at each level is large enough, then the probability of having
erroneous candidates can be made very small.

\bthe{ubound}
Suppose that (A1) and (A2) in \rlem{ubound} hold.
For any constant $C >0$, suppose that 
\beq{ubound3300}
\uu_{\MM-1} \ge \frac{1}{2} \log_2 \Big(\frac{K}{C}\Big),
\eeq
and 
for $m=\MM-2,\ldots, 0$,
\beq{ubound3333}
\uu_m \ge \frac{1}{2^{\MM-m}}\log_2 \Big(\frac{(1+C)^2}{C}K \Big).
\eeq
Then
$\ex [L_{\rm err}(m,\ell)] \le CK$ and $\ex [L(m,\ell)] \le (1+C)K$
for all $m, \ell$, and
\beq{ubound4444}
\pr (L_{\rm err}(0,0)\ge 1 ) \le (\frac{1}{2})^{\uu_0 2^{\MM}} (1+C)^2K^2.
\eeq
Thus, the probability that the $K$ codewords can be recovered correctly is at least
$$1-(\frac{1}{2})^{\uu_0 2^{\MM}} (1+C)^2K^2.$$
\ethe

\bproof
We first show by induction that
$\ex [L_{\rm err}(m,\ell)] \le CK$ and $\ex [L(m,\ell)] \le (1+C)K$
for all $m, \ell$ under \req{ubound3300} and \req{ubound3333}.
For $m=\MM-1$, it follow from \rlem{ubound}(i), \rlem{ubound}(ii), and \req{ubound3300}
that $\ex [L_{\rm err}(\MM-1,\ell)] \le CK$ and $\ex [L(\MM-1,\ell)] \le (1+C)K$.
Assume that $\ex [L_{\rm err}(m+1,\ell)] \le CK$ and $\ex [L(m+1,\ell)] \le (1+C)K$ hold for all $\ell=0,1,\ldots, 2^{m+1}-1$ as the induction hypotheses.
As a direct result of \rlem{ubound}(iii) and the induction hypotheses, we have
$\ex [L_{\rm err}(m,\ell)] \le CK$ and $\ex [L(m,\ell)] \le (1+C)K$ for all $\ell=0,1,\ldots, 2^{m}-1$.
Using these in \req{ubound1111} with $m=0$ yields
$$L_{\rm err}(0,0) \le (\frac{1}{2})^{\uu_0 2^{\MM}} (1+C)^2K^2.$$
The upper bound in \req{ubound4444} then follows from Markov's inequality.
\eproof

As $\ex [L(m,\ell)] \le (1+C)K$ for all $m, \ell$, the expected number of hash operations (for consistency check) for  merging two subtrees is bounded by
$(1+C)^2K^2$. Since there are $2^\MM-1$ non-leaf nodes (that needs to be merged) in a complete binary tree with $2^\MM$ leaf nodes, the expected total number of hash operations is bounded by $(2^\MM-1)(1+C)^2K^2$ if the parity bits are allocated to satisfy \req{ubound3300} and \req{ubound3333}.

Suppose that we choose $\uu_{\MM-1}=\lceil \frac{1}{2} \log_2 \Big(\frac{K}{C}\Big) \rceil \le \lceil \frac{1}{2} \log_2 \Big(\frac{(1+C)^2}{C}K\Big) \rceil $
and $\uu_m =\lceil \frac{1}{2^{\MM-m}}\log_2 \Big(\frac{(1+C)^2}{C}K \Big) \rceil$, $m=\MM-2,\ldots, 0$.
Then it follows from \rthe{ubound} that the probability of correct recovery of the $K$ codewords approaches to 1 as $n \to \infty$.
Using $\lceil x \rceil \le 1+x$, it is straightforward to show that the total number of parity bits in a sub-block
$$\sum_{m=0}^{\MM-1} \uu_m \le M+\log_2 K+ \log_2 \Big(\frac{(1+C)^2}{C}\Big).$$
Since there are $K$ codewords with $n=2^M$ sub-blocks, the total number of parity bits needed in our hierarchical stitching algorithm is
$$nK \Big( \log_2n+\log_2 K+ \log_2 \Big(\frac{(1+C)^2}{C}\Big)\Big).$$
This is quite close to the lower bound in \rprop{lbound}.

Note from \rthe{ubound} that for $K=300$ and $\MM=5$, there are $n=32$ sub-blocks. For $C=9$, we can choose $\uu_4=3$, $\uu_3=3$, and $\uu_2=2$ and $\uu_1=\uu_0=1$ to satisfy the condition in \req{ubound3300} and \req{ubound3333}. The total number of parity bits in each sub-block is 10 for this case, and the total number of parity bits in each codeword (with 32 sub-blocks) is 320 bits.
From \rthe{ubound}, the probability that the $300$ codewords can be recovered correctly (up to a permutation) is at least
$$1-(10)^2 (300)^2 (\frac{1}{2})^{32} \approx 0.998.$$

\end{document}